\documentclass[12 pt]{article}

\usepackage{amsmath,amssymb,amsthm,graphicx,latexsym}

\newcommand{\ed}{\end{document}}
\newcommand{\beq}{\begin{equation}}
\newcommand{\eeq}{\end{equation}}
\newcommand{\beqa}{\begin{eqnarray}}
\newcommand{\eeqa}{\end{eqnarray}}
\newcommand{\bc}{\begin{center}}
\newcommand{\ec}{\end{center}}
\newcommand{\vs}{\vspace}
\newcommand{\ba}{\begin{array}}
\newcommand{\ea}{\end{array}}

\begin{document}
\begin{center}
\bf{\Large{Conformal Invariance in Noncommutative Geometry and Mutually Interacting Snyder Particles}}\\
\end{center}
\vs{0.1cm}
\begin{center}
Souvik Pramanik$^*$ \footnote{E-mail: souvick.in@gmail.com},
Subir Ghosh$^*$ \footnote{E-mail: subir$_-$ghosh2@rediffmail.com}
Probir Pal$^\dagger$  \footnote{E-mail: probir.kumarpal@gmail.com}\\
\vspace{0.2cm}
\small{$^*$ \emph{Physics and Applied Mathematics Unit,\\ Indian Statistical Institute, 203 B. T. Road, Kolkata 700108, India} \\
\vspace{0.1cm}
$^\dagger$ \emph{Barasat Government College, 10 KNC Road, Barasat, Kolkata 700124, India}}
\end{center}
\vs{0.1cm}
\begin{abstract}
A system of relativistic Snyder particles with mutual two-body interaction that lives in a Non-Commutative Snyder geometry is studied. The underlying novel symplectic structure is a coupled and extended version of (single particle) Snyder algebra. In a recent work by Casalbuoni and Gomis, {\it{Phys.Rev. D90, 026001 (2014)}} \cite{cas}, a system of interacting conventional particles (in commutative spacetime) was studied with special emphasis on it's Conformal Invariance. Proceeding along the same lines \cite{cas} we have shown that our interacting Snyder particle model is also conformally invariant. Moreover, the conformal Killing vectors have been constructed. Our main emphasis is on the Hamiltonian analysis of the conformal symmetry generators. We demonstrate that the Lorentz algebra remains undeformed but validity of the full conformal algebra requires further restrictions.
\end{abstract}
\vskip .3cm

\section{Introduction}

Conformal invariance and Non-commutative geometry are two of the very active areas in modern physics. In the present work, we have united these two diverse themes in a specific conformally invariant model of mutually interacting relativistic particles, living in a Snyder form of Non-commutative geometry and  have studied its consequences. In particular, we extend the recent work of Casalbuoni and Gomis \cite{cas} to Non-commutative spacetime.

It is conventional to treat point particle mechanics as a field theory, compactified to one dimension. Indeed, this interpretation is not only of purely academic interest. In many cases it helps to understand particle excitation properties in the corresponding field theory, which is a higher dimensional extension of the discrete particle model. Example of such an interpretation is the connection between relativistic point particle (Einstein mass-energy relation) and Klein-Gordon scalar field theory (Klein-Gordon equation in momentum space). On the opposite side of the coin, there is the mapping between a linear chain of coupled oscillators that in the continuum limit reproduces the elastic wave model \cite{gold}, (or its higher derivative extension depending on the nature of the coupling in the discrete model \cite{sg}). Another advantage is that discrete mechanical models can be studied in more exhaustive details, (such as, the symmetries, explicit dynamics in presence of interactions, etc.),  than their field theoretic generalizations. Interestingly, the former  provides valuable information about the latter. With this motivation in mind there are a lot of recent activities in constructing  multi-particle  models with specific forms of interactions that respect specific symmetries  of physical relevance. A topical example is the interacting relativistic particle system with conformal invariance, as studied in \cite{cas}. A precursor to this, is the non-relativistic  one dimensional Calogero-Moser rational model \cite{cal} of a system of interacting particles, with two-body interactions.

The importance of conformal invariance in all branches of modern physics can not be overstressed. In order to explain scaling behavior of deep inelastic scattering in SLAC experiments \cite{bjor}, if one starts from the scale invariance in critical phenomena \cite{kad,wil}, then this  determines Operator-Product expansions as discovered  by Polyakov \cite{pol} and by Wilson \cite{wil1}. Later on the exciting AdS/CFT correspondence \cite{mal} helps to study strongly coupled boundary conformal field theories (identified with High $T_c$ superconductors \cite{gub}) from corresponding weakly coupled gravity theories. More relevant to the present work are the higher spin models, studied by Vasiliev \cite{vas}, based on massless Klein-Gordon equation. In the relativistic massless particle model, these are connected to conformal symmetry of the action. (See the Introduction section of \cite{cas} for a brief review and references.)

Another intensely active area of research in physics is Non-Commutative (NC) spacetime. It originated in the work of Snyder \cite{sn}, who introduced NC spacetime to ameliorate the ultraviolet singularity of Quantum Field Theory by introducing a short distance cutoff scale via NC spacetime. Unfortunately, this extension did not become popular. Surprisingly, in the work of Seiberg and Witten \cite{sw}, NC spacetime reappeared in a specific low energy limit of open strings with endpoints fixed on $D$-brans. Again, other {\it{avatars}} of NC spacetime have become very popular in predicting and analyzing Quantum Gravity signatures in low energy phenomena \cite{das}, since these new forms of NC spacetimes \cite{kemp} are compatible with Generalized Uncertainty Principle, that is advocated by high energy scattering amplitudes in String Theory \cite{ven}. It should be noted that there is a basic qualitative difference between the structure of the NC spacetime that appeared in \cite{sw} and \cite{sn,das,kemp}. The NC extension of the canonical Poisson brackets between configuration space or phase space variables are constant $c$-numbers in the former \cite{sw}, whereas they are non-constant and in particular operatorial in nature in the latter \cite{sn,das, kemp}. For this reason, in particle and field theories based on the former NC spacetime, fundamental spacetime symmetries are violated \cite{ncs}, whereas they can be kept intact in models based on the latter form of NC spacetimes \cite{sg1}. In this context, we have stressed in \cite{sgs} that it is essential to study the dynamics of these generalized particles in presence of interactions. There the first steps were taken in this direction, when generalized particle models were  studied in presence of gauge interactions \cite{sgs}. Also the thermodynamics of a gas of non-interacting generalized particles was analyzed in \cite{sud}.

In the present work, we will study in detail the symmetries concerning a conformally invariant, relativistic, non-commutative and mutually interacting massless two particle system. Massless particles are considered  since a mass parameter will invariably brake conformal invariance. The motivation behind this project is to construct a multi-particle system having the above features. We will closely follow the approach of \cite{cas}. The explicit NC relativistic particle model, that we use, is close in spirit to a free {\footnote{This particular NC extension \cite{rb} of the conventional massive free particle action  introduces primary constraints in the system and a Hamiltonian constraint analysis along Dirac's formulation \cite{dir} revealed that the symplectic structure of the NC particle obeyed the Snyder algebra \cite{sn}. However, in \cite{rb} only the "free" Snyder particle was considered: free in the sense that it consisted a single particle phase space degrees of freedom.}} {\it{massive}} Snyder particle model proposed by Banerjee, Kulkarni and Samanta \cite{rb}. However, there is a subtlety involved since the massless version of the massive Snyder model of \cite{rb} does not satisfy the Snyder algebra. This will be explained later. But in our model the interaction saves the day, since it generates an effective (non-constant) mass term so that a generalized Snyder algebra reappears. Hence, in our work we have non-trivially extended the free massive model of \cite{rb} to a massless two (Snyder) particle system with a mutual inter-particle interaction. We reveal interesting and qualitative differences between our interacting massless two-particle Snyder model and the massive single particle Snyder model \cite{rb}.

We have analyzed the symmetry structure in both from Lagrangian and Hamiltonian perspectives, in a two-particle system. Indeed, the Lagranian model can be extended in a straightforward way to a multi-particle system. Explicit computations of the many particle symplectic structure will be worth pursuing which we postpone for a future publication.  In the Lagrangian framework,  conformal Killing vectors are constructed and the general analysis essentially follows \cite{cas} albeit with important differences.

Our analysis in the Hamiltonian framework is completely new, where we, for the  first time, provide a novel form of coupled Snyder algebra. In our two particle model, the constraint analysis is quite involved with many non-vanishing brackets, emerging as a result of the mutual interaction. But most interestingly, {\it{we establish the presence of a First Class Constraint (FCC)}} (see \cite{dir} and our analysis below), {\it{ showing that the system has a gauge invariance.}} This observation is quite in contrast with the free massless and massive model of \cite{rb} that have two FCCs and no FCC respectively. On the other hand our result is similar to the presence of a gauge invariance in the interacting two-particle conventional model \cite{cas}. Finally, from the extremely involved two-particle Snyder algebra, we have been able to explicitly construct Lorentz generators for the two particle system, $J^{\mu\nu}= J^{\mu\nu}_1+J^{\mu\nu}_2$, where $J^{\mu\nu}_i$ consist of the $i$'th particle phase space coordinates. Incidentally, it is worthwhile to point out that, $J^{\mu\nu}$ transforms each coordinate and momentum variable canonically and satisfies the undeformed Lorentz algebra, whereas, $J^{\mu\nu}_i$ do not correctly transform any of the degrees of freedom and do not separately satisfy a Lorentz algebra. We find surprising results regarding rest of the generators, i.e. for Translation $T^\mu $, Dilatation $D$ and Special Conformal Transformation generators $K^\mu $ which are constructed as $T^\mu=T^\mu _1 +T^\mu _2$, $D=D_1 +D_2$ and $K^\mu =K^\mu _1 +K^\mu _2$, with the subscript index denoting the first and second particle degrees of freedom. To recover correct transformation and the algebra regarding $T^\mu $, we find that a "rigidity" condition in the momentum sector (to be explained later) comes in to play, to generate correct transformations and algebra. This can be understood from the specific nature of interaction that depends only on the relative coordinates. However, $K^\mu $ requires still another restriction in the coordinate sector, that tends to reduce the system to a single particle one with a self interaction.  The present analysis indicates that, constrained systems can show qualitatively different characteristics in Lagrangian and Hamiltonian framework, even in classical analysis (that is not taking in to account operator ordering and other quantum features). It seems that, all symmetries of the Lagrangian are not strictly obeyed in the Hamiltonian framework and some of the (Lagrangian) symmetries are satisfied in a hypersurface of the full phase space, leading to restricted invariance in the Hamiltonian setup. This can have interesting consequences upon quantization, and is reminiscent of the well known quantum anomaly phenomena, where not all classical symmetries are maintained upon quantization.

The paper is organized as follows: in Section 2, we introduce the interacting two particle  Snyder model action. In Section 3, the Lagrangian analysis of the symmetry is given based on Killing vectors. The detailed constraint analysis and the resulting Dirac bracket algebra are derived in Section 4. In Section 5, we construct explicitly the Conformal Symmetry generators and demonstrate that the generators satisfy the undeformed Conformal algebra (subject to additional restrictions).  Our conclusions and future prospects of the present work are listed in  Section 6.

\section{Interacting two-particle Snyder model}
We start our discussion with the massive and massless single Snyder particle model \cite{rb}.
\subsection{Massive and massless single particle Snyder model:}
At the outset, let us recall the required symmetry transformations that our system has to satisfy. The action should be reparametrization invariant, as well as invariant under dilatations. It can be shown that our chosen model is invariant under the dilatation $x_\mu\rightarrow \lambda x_\mu$, $e\rightarrow\lambda^2 e$, where $e$ is the einbein in Polyakov formulation. In addition,  here we are interested in special conformal symmetry. Since, the special conformal transformation can be obtained through the series of operations: (inversion)$\bigotimes$(translation)$\bigotimes$(inversion),  as discussed in \cite{cas, kas}, the inversion operation
\begin{equation}
x_\mu \rightarrow \frac{x_\mu}{x^2}
\label{20}
\end{equation}
plays a crucial role in the present context. This means our model needs to be invariant under inversion as well.

In order to show how the Snyder algebra is induced via constraints, let us start with the free massive Snyder particle \cite{rb}. The Nambu-Goto action \cite{rb}, with our shorthand notation $(ab)=a^\mu b_\mu=\eta^{\mu\nu}a_\mu b_\nu$, $diag~\eta^{00}=-\eta^{ii} =1$, becomes
\begin{equation}
S=-m\int~d\tau {\sqrt{\dot{x}^2-\frac{(x\dot x)^2}{x^2}}}=\int L~d\tau.
\label{ng1}
\end{equation}
The canonical momenta are
\begin{equation}
p_\mu = -\frac{\partial L}{\partial \dot{x}^\mu}\frac{m}{\sqrt{\dot{x}^2-\frac{(x\dot{x})^2}{x^2}}}\left(\dot x_\mu -\frac{(x\dot x)x_\mu}{x^2}\right).
\label{ng2}
\end{equation}
They give rise to two constraints:
\begin{equation}
\chi_1\equiv p^2-m^2\approx 0,~~\chi_2\equiv (px) \approx 0.
\label{ng3}
\end{equation}
Using the canonical Poisson brackets $\{x_\mu ,p_\nu \}=-\eta_{\mu\nu},$ $\{x_\mu ,x_\nu \}=\{p_\mu ,p_\nu \}=0$, it can be shown that these two constraints do not commute (in the sense of Poisson brackets), and therefore are characterized as Second Class Constraints (SCC) \cite{dir},
\begin{equation}
\{\chi_1,\chi_2\}=2p^2=2\chi_1+ 2 m^2\approx 2 m^2,~~\{\chi_1,\chi_1\}=\{\chi_2,\chi_2\}=0.
\label{ng3}
\end{equation}

In the terminology of Dirac constraint analysis \cite{dir}, the
noncommutating constraints are termed as SCC and the commutating constraints, that induces local gauge invariance, are named First Class Constraints (FCC). In a generic Second Class system with $n$  SCCs $\chi_i$, $i=1,2,..n$, the modified symplectic structure (or Dirac brackets) are defined in the following way,
\begin{equation}
\{A,B\}^*=\{A,B\}-\{A,\chi _i\}\{\chi ^i,\chi ^j\}^{-1}\{\chi _j,B\}, \label{a6}
\end{equation}
where $\{\chi ^i,\chi ^j\}$ is the invertible constraint matrix. From now on we will use $\{,\}$ notation instead of $\{,\}^*$ for Dirac brackets.

For the above set off SCCs (\ref{ng3}) of the Snyder model, the resulting Dirac bracket algebra,
\begin{equation}
\{x_\mu ,x_\nu \}=\frac{1}{m^2}(x_\mu p_\nu-x_\nu p_\mu)~,~~\{x_\mu,p_\nu\}=-\delta_{\mu\nu}+\frac{1}{m^2}p_\mu p_\nu~,~~\{p_\mu,p_\nu\}=0,
\label{dir}
\end{equation}
is identical to the celebrated Snyder algebra \cite{sn}. Note that $1/m^2$ plays the role of the NC parameter. It is at once clear that the massless limit of this algebra is singular. In fact the massless limit of the Nambu-Goto form of particle action itself does not exist.

Since our interest lies in the massless particles, we start from the Polyakov version of the massive action by introducing the einbein $e$ to obtain the equivalent massive Snyder particle action,
\begin{equation}
S=-\int \frac{1}{2}\left[\frac{1}{e} \left(\dot{x}^2-\frac{(x\dot{x})^2}{x^2}\right)+ m e \right]~d\tau.
\label{}
\end{equation}
Thereafter we take the massless limit,
\begin{equation}
S=-\int \frac{1}{2 e} \left(\dot{x}^2-\frac{(x\dot{x})^2}{x^2}\right)~d\tau.
\label{ml}
\end{equation}
Let us check the invariance of (\ref{ml}) under inversion $x^\mu \rightarrow \frac{x^\mu}{x^2}$. With this transformation the part $\left(\dot{x}^2-\frac{(x\dot{x})^2}{x^2}\right)$ $\rightarrow$ $\frac{1}{(x^2)^2}\times \left(\dot{x}^2-\frac{(x\dot{x})^2}{x^2}\right)$. Since our Lagrangian is of the form $L=-\frac{1}{2 e}\left(\dot{x}^2-\frac{(x\dot{x})^2}{x^2}\right)$,  it will be invariant under the above inversion if $e$ is transformed as $e\rightarrow \frac{e}{(x^2)^2}$. Notice that this transformation rule is same as that of a commutative free particle \cite{cas}.

Now varying the above action with respect to the einbein $e$ produces
\begin{equation}
\frac{1}{2 e^2}\left(\dot{x}^2-\frac{(x\dot{x})^2}{x^2}\right)=0~\Rightarrow ~~\dot{x}^2-\frac{(x\dot{x})^2}{x^2}=0.
\label{m0}
\end{equation}
On the other hand, since $p_\mu =-\frac{\partial L}{\partial \dot{x}^\mu}=\frac{1}{e}\left(\dot x_\mu -\frac{(x\dot x)x_\mu}{x^2}\right)$, using (\ref{m0}) we obtain
\begin{equation}
p^2=\frac{1}{e^2}\left(\dot{x}^2-\frac{(x\dot{x})^2}{x^2}\right)=0~,~~(xp)=0.\nonumber
\label{}
\end{equation}
Therefore there are two constraints $\psi_1\equiv p^2\approx0$, the dispersion relation for a massless particle, and $\psi_2\equiv(xp)\approx0$. However, now the two constraints commute:
\begin{equation}
\{\psi_1,\psi_2\}=2p^2=2\psi_1\approx 0;~~\{\psi_1,\psi_1\}=\{\psi_2,\psi_2\}=0.
\label{}
\end{equation}
Hence there does not appear any non-commuting SCCs and both the constraints are FCC in this massless case as we have mentioned earlier. This shows that  the massless version of the massive Snyder model of \cite{rb} does not live in a Snyder phase space. We will now show how an interaction cures this problem.
\subsection{Massless two particle interacting Snyder model:} After all these preliminaries we are finally ready to introduce our model of two massless interacting Snyder particles:
\begin{equation}
S=-\int \left[\frac{1}{2 e_1}\left(\dot{x}_1^2-\frac{(x_1\dot{x}_1)^2}{x_1^2}\right)+\frac{1}{2 e_2}\left(\dot{x}_2^2-\frac{(x_2\dot{x}_2)^2}{x_2^2}\right)+\frac{\alpha^2}{4}\frac{\sqrt{e_1 e_2}}{r^2}\right] d\tau
\label{21}
\end{equation}
where $r_{\mu}=x_{1\mu}-x_{2\mu}$ and $\{x^\mu,p^\nu\}=-\eta^{\mu\nu}$, $\{e_i,\pi_j\}=\delta_{i j}$. First of all, note that for the relative coordinate $r_\mu$,  the inversion transformation reads $r^2\rightarrow \frac{r^2}{(x_1^2)(x_2^2)}$. Therefore, a conformally invariant interaction term between the two particle can be chosen as $\frac{\sqrt{e_1 e_2}}{r^2}$. Incidentally, the interaction term is same as that in \cite{cas} for the commutative interacting particle.  The invariance of the kinetic terms under  inversion transformation has already been been discussed in  Section 2.1.

The momenta corresponding to the Lagrangian (\ref{21}) are $p_{i}^\mu =-\frac{\partial L}{\partial \dot x_{i \mu} }=\frac{1}{e_i}\left(\dot x_{i\mu} -\frac{(x_i \dot{x}_i)x_{i\mu}}{x_i^2}\right)$, $\pi_{i}=\frac{\partial L}{\partial \dot e_i}=0$
leading to four primary constraints,
\begin{equation}
 \pi_1 \approx 0,~ \pi_2 \approx 0,~ (x_1 p_1)\approx 0,~(x_2 p_2)\approx 0.
\label{}
\end{equation}
Therefore the extended Hamiltonian can be written as
\begin{equation}
H_D=-\frac{e_1}{2}p_1^2-\frac{e_2}{2}p_2^2+\frac{\alpha^2}{4}\frac{\sqrt{e_1 e_2}}{r^2}+\lambda_1 \pi_1+\lambda_2 \pi_2 + \lambda_3(x_1 p_1)+\lambda_4(x_2 p_2),\label{21h}
\end{equation}
where $\lambda_i$ are Lagrange multipliers that incorporate the primary constraints. Time persistence of the primary constraints yields the following secondary constraints,
\begin{eqnarray}
\Phi_1 &=& \frac{1}{2}\left(p_1^2-\frac{\alpha^2}{4}\sqrt{\frac{e_2}{e_1}}\frac{1}{r^2}\right)\approx 0~,~~
\Phi_2 =  \frac{1}{2}\left(p_2^2-\frac{\alpha^2}{4}\sqrt{\frac{e_1}{e_2}}\frac{1}{r^2}\right)\approx 0\nonumber\\
\Phi_3 &=& e_1 p_1^2-\frac{\alpha^2}{2}\frac{\sqrt{e_1 e_2}}{r^4}(x_1.r)\approx 0~,~~
\Phi_4 = e_2 p_2^2+\frac{\alpha^2}{2}\frac{\sqrt{e_1 e_2}}{r^4}(x_2.r)\approx 0
\end{eqnarray}
However, working with such a large number (eight) of constraints is extremely unwieldy and we take recourse to a much more compact form of the Lagrangian by eliminating the $e_i$.

By varying the action (\ref{21}) with respect to the einbeins $e_1$, $e_2$ and subsequently by solving we get
\begin{equation}
\frac{1}{e_1} =\frac{\alpha \left[\left(\dot{x}_1^2-\frac{(x_1 \dot{x}_1)^2}{x_1^2}\right)\left(\dot{x}_2^2-\frac{(x_2 \dot{x}_2)^2}{x_2^2}\right)\right]^\frac{1}{4}}{2 \sqrt{r^2}\left(\dot{x}_1^2-\frac{(x_1 \dot{x}_1)^2}{x_1^2}\right)}~,~~
\frac{1}{e_2} =\frac{\alpha \left[\left(\dot{x}_1^2-\frac{(x_1 \dot{x}_1)^2}{x_1^2}\right)\left(\dot{x}_2^2-\frac{(x_2 \dot{x}_2)^2}{x_2^2}\right)\right]^\frac{1}{4}}{2 \sqrt{r^2}\left(\dot{x}_2^2-\frac{(x_2 \dot{x}_2)^2}{x_2^2}\right)},
\label{ee}
\end{equation}
where $\alpha \ne 0$. Substituting these two expression into the above action we obtain
\begin{equation}
S=-\alpha \int \left[\frac{\left(\dot{x}_1^2-\frac{(x_1 \dot{x}_1)^2}{x_1^2}\right)\left(\dot{x}_2^2-\frac{(x_2 \dot{x}_2)^2}{x_2^2}\right)}{r^4}\right]^\frac{1}{4} d\tau.
\label{ac}
\end{equation}
This is the form of the action that we will study from now on. It needs to be mentioned that the $\alpha \rightarrow 0$ limit is not smooth since, following (\ref{ee}), in this limit $e_1,e_2$ becomes singular. Hence the Dirac bracket algebra derived later in Section 4 will not reduce to two decoupled Snyder algebra in the limit $\alpha \rightarrow 0$.

Notice that the structure of the action is close to the analogous one in \cite{cas}. The action vanishes for $\alpha =0$ showing that the free two-particle  conformal invariant action vanishes. This point was discussed in \cite{cas}. But in the present work the condition $\alpha \ne0$ assumes much more significance since a vanishing $\alpha$ will make the Dirac bracket algebra (computed later) singular. Comparing with the massive single Snyder particle Lagrangian (\ref{ng1}), we notice that the structure of the massive model has reappeared in the interacting massless model.

\section{Analysis of the rigid symmetries and conformal Killing vectors}
The free Snyder particle is conformally invariant as we have shown in Section 2 and it's symmetry group is $SO(D,2)$. Quite obviously for a system of two non-interacting  Snyder particles the group is extended to $SO(D,2)_1\otimes SO(D,2)_2$ that act on particle $1$ and $2$ respectively. Once a conformally invariant interaction is introduced, the group reduces to its diagonal subgroup. In this section we discuss how this is reflected on the associated conformal Killing vectors. Our analysis essentially follows \cite{cas}. In the next section we will analyze the symmetry in a Hamiltonian framework.

The conserved quantities are defined in terms of Killing vectors as
\begin{equation}
 G=\sum_{i=1}^2\xi_{i\mu}(x_1,x_2)p_i^\mu.\label{eq:5.1}
\end{equation}

In the above $\xi_{i\mu}$ are the Killing vectors and from (\ref{ac}) $p_i^{\mu} $ denote the momenta of the $i^{th}$ particle,
\begin{equation}
p_i^\mu=-\frac{\partial L}{\partial \dot x_{i\mu}}=\frac 12 \frac{\left(\dot x_i^\mu-\frac{(x_i \dot{x}_i)}{x_i^2}x_i^\mu\right)}{\left(\dot x_i^2-\frac{(x_i \dot{x}_i)^2}{x_i^2}\right)}L .\label{eq:5.2}
\end{equation}
From (\ref{ac}) the Lagrangian equations of motion we have,
\begin{eqnarray}
\dot p_1^\mu &=& -\frac{d}{d t}\left(\frac{\partial L}{\partial \dot{x}_{1 \mu}}\right)=-\frac{\partial L}{\partial x_{1\mu}}=-\frac{1}{2}\frac{\left(\dot x_1^\mu-\frac{(x_1 \dot{x}_1)}{x_1^2}x_1^\mu\right)}{\left(\dot x_1^2-\frac{(x_1 \dot{x}_1)^2}{x_1^2}\right)}\frac{(x_1.\dot{x}_1)L}{x_1^2}-\frac 12 \frac{r_\mu}{r^2}L,\nonumber\\
\dot p_2^\mu &=& -\frac{d}{d t}\left(\frac{\partial L}{\partial \dot{x}_{2 \mu}}\right)=-\frac{\partial L}{\partial x_{2\mu}}=-\frac{1}{2}\frac{\left(\dot x_2^\mu-\frac{(x_2 \dot{x}_2)}{x_2^2}x_2^\mu\right)}{\left(\dot x_2^2-\frac{(x_2 \dot{x}_2)^2}{x_2^2}\right)}\frac{(x_2.\dot{x}_2)L}{x_2^2}+\frac 12 \frac{r_\mu}{r^2}L.\label{p-dot}
 \end{eqnarray}
From the requirement of $G$ to be a conserved quantity $\dot G=0$, its time derivative vanishes. Therefore taking the time derivative of $G$, using the expression of the momenta  given in (\ref{eq:5.2}) and the Lagrange equations of motion (\ref{p-dot}), we obtain
\begin{eqnarray}
 \label{dotg}
\dot G &=& \frac 12\sum_{i,j=1}^2\left(\partial_{j\mu}\xi_{i\nu}(x_1,x_2)\right)\frac{\dot{x}_j^\mu\left(\dot x_i^\nu-\frac{(x_i \dot{x}_i)}{x_i^2}x_i^\nu\right)}{\left(\dot x_i^2-\frac{(x_i \dot{x}_i)^2}{x_i^2}\right)}L-\frac{1}{2}\sum_{i=1}^2\xi_{i\mu}(x_1,x_2)\frac{\left(\dot x_i^\mu-\frac{(x_i \dot{x}_i)}{x_i^2}x_i^\mu\right)}{\left(\dot x_i^2-\frac{(x_i \dot{x}_i)^2}{x_i^2}\right)}\frac{(x_i\dot{x}_i)}{x_i^2} L\nonumber\\
&&-(\xi_{1\mu}(x_1,x_2)-\xi_{2\mu}(x_1,x_2))\frac{r^\mu L}{r^2}=0,
\end{eqnarray}
where $\partial_{i\mu}={\partial}/{\partial x_i^{\mu}}$. Now a necessary condition to have a solution is $\partial_{j\mu}\xi_{i\nu}(x_1,x_2)\mid_{j\neq i}=0$ $\Rightarrow \xi_{i\nu}=\xi_{i\nu}(x_i)$. If we substitute this condition in (\ref{dotg}) we get
\begin{eqnarray}
&&\frac 12\sum_{i=1}^2\left(\partial_{i\mu}\xi_{i\nu}(x_i)\right)\frac{\dot{x}_i^\mu\left(\dot x_i^\nu-\frac{(x_i \dot{x}_i)}{x_i^2}x_i^\nu\right)}{\left(\dot x_i^2-\frac{(x_i \dot{x}_i)^2}{x_i^2}\right)}L-\frac{1}{2}\sum_{i=1}^2\xi_{i\mu}(x_i)\frac{\left(\dot x_i^\mu-\frac{(x_i \dot{x}_i)}{x_i^2}x_i^\mu\right)}{\left(\dot x_i^2-\frac{(x_i \dot{x}_i)^2}{x_i^2}\right)}\frac{(x_i\dot{x}_i)}{x_i^2} L\nonumber\\
&&-(\xi_{1\mu}(x_1)-\xi_{2\mu}(x_2))\frac{r^\mu L}{r^2}=0.\label{kill-1}
\end{eqnarray}
Notice that a solution of (\ref{kill-1}) is possible provided the explicit $\dot x_i^\mu $ dependence are removed. In the first term this is possible if $\partial _{i\mu }\xi _{i\nu }$ has a symmetric form $\sim\eta_{\mu\nu }$ so that the $\dot x_i^\mu $-dependence cancels out. In  a similar calculation in \cite{cas}, a condition of this type was enough to remove the $\dot x_i^\mu $-dependence completely. But in our case another condition is needed to remove  $\dot x_i^\mu $ dependence from the second term. That condition is  $\xi_{i\mu}(x_i)=\lambda_i(x_i)x_{i\mu}$  for which the second term vanishes. Note that this is consistent with the previous condition $\partial _{i\mu }\xi _{i\nu }\sim\eta_{\mu\nu }$. Thus the $\dot x$-dependence is removed  and the solution of the equation (\ref{kill-1}) can be written as
\begin{eqnarray}
\partial_{i\mu}\xi_{i\nu}(x_i)&=& \eta_{\mu\nu}\lambda_{i}(x_i)+x_{i\nu}\partial_{i\mu}\lambda_i(x_i),~~~i=1,2,\nonumber\\
\frac 12 \sum_{i=1}^2\,\lambda_{(i)} &=& {(\xi_{1\mu}-\xi_{2\mu})r^\mu}\frac 1{r^2} \label{sol-1}
\end{eqnarray}
Now $\xi_i$ will satisfy the  conformal Killing equation if the last term of the first equation of (\ref{sol-1}) vanishes, i.e. $\partial_{i\mu}\lambda_i(x_i)=0$. This implies $\lambda_i(x_i)=c_i$, where $c_i$'s are constant. Therefore the solutions reduce to
\begin{eqnarray}
\frac{1}{2} \left(\partial_{i\mu}\xi_{i\nu}(x_i)+\partial_{i\nu}\xi_{i\mu}(x_i)\right)&=& \eta_{\mu\nu}\lambda_{i}(x_i),~~~i=1,2,\nonumber\\
\frac 12 \sum_{i=1}^2\,\lambda_{(i)} &=& {(\xi_{1\mu}-\xi_{2\mu})r^\mu}\frac 1{r^2}, \label{sol-2}
\end{eqnarray}
where we have used the symmetric part of $\partial_{i\mu}\xi_{i\nu}$. Thus there appears to be two independent conformal Killing vectors $\xi_{1\mu}(x_1)=c_1 x_
{1\mu}$ and $\xi_{2\mu}(x_2)=c_2 x_
{2\mu}$ associated with the two conformal groups $SO(D,2)_i$ acting on the two variables $x_1$ and $x_2$ respectively. These are the symmetry groups of two massless non-interacting particles. But from the second equation of (\ref{sol-2}) we have $\frac 12 (c_1+c_2)=\frac{(c_1 x_{1\mu}-c_2 x_{2\mu})r^\mu}{r^2}$, which is satisfied if and only if $c_1=c_2$. Therefore we have $\lambda_1(x_1)=\lambda_2(x_2)=c$. Hence $\xi_{1\mu}(x_1)=c x_
{1\mu}$ and $\xi_{2\mu}(x_2)=c x_
{2\mu}$ are the conformal Killing vectors of our system. Thus the second condition (\ref{sol-2}) is satisfied if and only if the infinitesimal parameters defining the two Killing vectors are identical. Thus the symmetry $SO(D,2)_1\otimes SO(D,2)_2$ reduces to the diagonal subgroup $SO(D,2)$ due to the interaction between the two particles. A similar conclusion has  been reached \cite{cas} for the interacting conventional particle model.

\section{Constraint analysis and coupled Snyder algebra}
In this section we start with the Lagrangian (\ref{ac}), from which the momenta $p_i^\mu=-\frac{\partial L}{\partial \dot{x}_{i \mu}}$, $i= 1, 2$, corresponding to the variable $\dot{x}_i^\mu$, are computed,
 $$
p_i^\mu=\frac{L}{2}\frac{\left(\dot{x}_i^\mu-\frac{(x_i\dot{x}_i)x_i^\mu}{x_i^2}\right)}{\left(\dot{x}_i^2-\frac{(x_i \dot{x}_i)^2}{x_i^2}\right)}.
$$
We immediately find two constraints,
\begin{equation}
(p_1x_1)=0~,~~(p_2x_2)=0.
\label{dd}
\end{equation}
Furthermore, from
$$ p_1^2 = \frac{\alpha^2}{4 r^2}
\frac{\left(\dot{x}_2^2-\left(\frac{x_2 \dot{x}_2}{x_2}\right)^2\right)^\frac{1}{2}}
{\left(\dot{x}_1^2-\left(\frac{x_1 \dot{x}_1}{x_1}\right)^2\right)^\frac{1}{2}}~,~~
p_2^2 = \frac{\alpha^2}{4 r^2}
\frac{\left(\dot{x}_1^2-\left(\frac{x_1 \dot{x}_1}{x_1}\right)^2\right)^\frac{1}{2}}
{\left(\dot{x}_2^2-\left(\frac{x_2 \dot{x}_2}{x_2}\right)^2\right)^\frac{1}{2}},$$
there appears another constraint
\begin{equation}
p_1^2 p_2^2-\frac{\alpha^4}{16 r^4}=0.
\label{cc}
\end{equation}
It is worthwhile to point out that (\ref{cc}) looks same as a  constraint arising in a similar way in \cite{cas}, but in our model we have additional constraints (\ref{dd}). We denote the constraints as
\begin{equation}
\chi_1\equiv (p_1x_1)\approx 0, ~\chi_2\equiv (p_2 x_2)\approx 0,~ \chi_3\equiv p_1^2 p_2^2-\frac{\alpha^4}{16 r^4}\approx 0.
\label{con}
\end{equation}
The following constraint algebra
\begin{eqnarray}
\{\chi_1,\chi_3\} &=& 2 p_1^2 p_2^2-\frac{\alpha^4(x_1r)}{4 r^6}~,~~
\{\chi_2,\chi_3\} = 2 p_1^2 p_2^2+\frac{\alpha^4(x_2r)}{4 r^6},\nonumber\\
\{\chi_1,\chi_1\} &=& \{\chi_1,\chi_2\}=\{\chi_2,\chi_2\}=\{\chi_3,\chi_3\}=0,\nonumber
\end{eqnarray}
reveals that $\{(\chi_1+\chi_2),\chi_3\}=4p_1^2 p_2^2-\frac{\alpha^4((x_1-x_2).r)}{4 r^6}=  4\chi_3$. Hence we have an FCC
\begin{equation}
D\equiv \chi_1+\chi_2 \approx 0.
\label{fcc}
\end{equation}
This is the First Class Constraint that was advertised earlier.
Later on we will show $D$ is identified as the Dilatation operator. The remaining pair of constraints
\begin{equation}
\chi_1-\chi_2 \approx 0,~~\chi_3\approx 0,
\label{scc}
\end{equation}
are SCC in nature, with the only non-vanishing bracket,
\begin{equation}
\{(\chi_1-\chi_2),\chi_3\}=-\frac{\alpha^4((x_1+x_2).r)}{4 r^6}\equiv \Lambda.
\label{scc}
\end{equation}
We choose a particular combination $\chi_1-\chi_2 \approx 0$ in order to maintain a symmetry between the two particle degrees of freedom. The final Dirac brackets of this system are,
\begin{eqnarray}
\{x_1^\mu,x_1^\nu\} &=& -\frac{2 p_2^2}{\Lambda}(x_1^\mu p_1^\nu-p_1^\mu x_1^\nu)~,~~
\{x_1^\mu,p_1^\nu\}= \eta^{\mu\nu}+\frac{\alpha^4}{4\Lambda r^6}x_1^\mu r^\nu-\frac{2 p_2^2}{\Lambda}p_1^\mu p_1^\nu~,\nonumber\\
\{p_1^\mu,p_1^\nu\} &=& -\frac{\alpha^4}{4\Lambda r^6}(p_1^\mu r^\nu-r^\mu p_1^\nu)~,\label{dir1}\\
\{x_2^\mu,x_2^\nu\} &=& \frac{2 p_1^2}{\Lambda}(x_2^\mu p_2^\nu-p_2^\mu x_2^\nu)~,~~
\{x_2^\mu,p_2^\nu\} = \eta^{\mu\nu}+\frac{\alpha^4}{4\Lambda r^6}x_2^\mu r^\nu+\frac{2 p_1^2}{\Lambda}p_2^\mu p_2^\nu~,\nonumber\\
\{p_2^\mu,p_2^\nu\} &=& -\frac{\alpha^4}{4\Lambda r^6}(p_2^\mu r^\nu-r^\mu p_2^\nu)~,\label{dir2}\\
\{x_1^\mu,x_2^\nu\} &=& -\frac{2}{\Lambda}(p_1^2 x_1^\mu p_2^\nu+p_2^2 p_1^\mu x_2^\nu)~,~~
\{p_1^\mu,p_2^\nu\}=\frac{\alpha^4}{4\Lambda r^6}(p_1^\mu r^\nu-r^\mu p_2^\nu)~,\nonumber\\
\{x_1^\mu,p_2^\nu\} &=& -\frac{\alpha^4}{4\Lambda r^6}x_1^\mu r^\nu+\frac{2 p_2^2}{\Lambda}p_1^\mu p_2^\nu~,~~
\{x_2^\mu,p_1^\nu\} =-\frac{\alpha^4}{4\Lambda r^6}x_2^\mu r^\nu-\frac{2 p_1^2}{\Lambda}p_2^\mu p_1^\nu.\label{dir3}
\end{eqnarray}
The above algebra has several interesting features.\\
(i) In the non-canonical terms that appear in the RHS of the brackets, the coordinates and momenta appear more symmetrically than the original single-particle Snyder algebra (\ref{dir}).\\
(ii) The momentum brackets, $\{p_1^\mu,p_1^\nu \}$, $\{p_2^\mu,p_2^\nu \}$ are non-zero, unlike the free Snyder case. Furthermore, presence of all the non vanishing cross-brackets between particle $1$ and particle $2$ make the system complicated to study.

The canonical Hamiltonian vanishes,
\begin{equation}
H=(p_1\dot x_1)+(p_2\dot x_2)-L=0,
\end{equation}
showing that the reparametrization invariance is intact and the dynamics will be generated by the FCC.

In the following section we will see how the above complicated Dirac Bracket structure can be compartmentalized so that the generators of the conformal invariance emerge in the Hamiltonian framework.
\section{Generators of Conformal invariance }
The symmetry group $SO(D,2)$ has $\frac{1}{2}(D+2)(D+1)$ number of generators which in our four-dimensional case is $15$. The full conformal algebra is given by
\begin{eqnarray}
\{J^{\mu\nu },J^{\alpha\beta }\} &=& g^{\mu\beta }J^{\nu\alpha
}+g^{\mu\alpha }J^{\beta \nu}+g^{\nu\beta }J^{\alpha\mu
}+g^{\nu\alpha }J^{\mu\beta }~,~~\{J^{\mu\nu },T^\sigma \}=g^{\nu\sigma}T^{\mu}-g^{\mu\sigma}T^{\nu}~,~~\nonumber\\
\{J^{\mu\nu },D \} &=& 0~,~~\{J^{\mu\nu },K^\sigma\}= 2D(g^{\nu\sigma}x^{\mu}-g^{\mu\sigma}x^{\nu})-x^2(g^{\nu\sigma}T^{\mu}-g^{\mu\sigma}T^{\nu})~,~~\{T^\mu ,T^\nu \}=0~,\nonumber\\
\{T^\mu ,D\} &=& T^\mu~,~~\{T^\mu ,K^\nu \}= 2Dg^{\mu\nu}-2J^{\mu\nu}~,~~\{D,D \}=0~,~~\{D,K^\mu \}=K^\mu ~,\nonumber\\
\{K^\mu ,K^\nu \}&=& 0, \label{0s}
\end{eqnarray}
where  $J^{\mu\nu}$, $T^\mu$, $D$ and $K^\mu$ stand for generators of Lorentz rotation, translation, dilatation and special conformal transformation respectively. Their explicit canonical expressions for a single particle degrees of freedom $x_\mu$, $p_\nu$ are given by,
\begin{equation}
J^{\mu\nu}=x^\mu p^\nu -x^\nu p^\mu~,~~T^\mu = p^\mu~,~~D=(xp)~,~~K^\mu =2(xp)x^\mu -x^2 p^\mu.\label{1s}
\end{equation}
Let us start with the most important one, the Lorentz generator $J^{\mu\nu}$. Since it is not clear how the interaction will modify the structure of the generators, we follow the results for the single Snyder particle model and taking cue from \cite{rb} we define $$J_1^{\mu\nu}=x_1^\mu p_1^\nu-x_1^\nu p_1^\mu~,~~J_2^{\mu\nu}=x_2^\mu p_2^\nu-x_2^\nu p_2^\mu.$$ Using the Dirac brackets (\ref{dir1},\ref{dir2},\ref{dir3}) we compute,
\begin{equation}
\{J_1^{\mu \nu},x_1^\sigma\}=(\eta^{\mu\sigma} x_1^\nu-\eta^{\nu\sigma} x_1^\mu)-\frac{\alpha^4 x_1^\sigma}{4 \Lambda r^6}(x_1^\mu r^\nu -x_1^\nu r^\mu),~\{J_2^{\mu \nu},x_1^\sigma\}=\frac{\alpha^4 x_1^\sigma}{4 \Lambda r^6}(x_2^\mu r^\nu -x_2^\nu r^\mu).
\label{jx}
\end{equation}
Indeed the coordinates do not transform correctly under $J_i^{\mu\nu}$ but now observe the miraculous cancelation to yield, (with a similar calculation for $x_2^\mu$ not shown here),
\begin{equation}
\{J^{\mu \nu},x_1^\sigma\}=\{J_1^{\mu \nu}+J_2^{\mu \nu},x_1^\sigma\}=(\eta^{\mu\sigma} x_1^\nu-\eta^{\nu\sigma} x_1^\mu)~,~~
\{J^{\mu \nu},x_2^\sigma\}=\{J_1^{\mu \nu}+J_2^{\mu \nu},x_2^\sigma\}=(\eta^{\mu\sigma} x_2^\nu-\eta^{\nu\sigma} x_2^\mu).
\label{jx1}
\end{equation}
Hence $x_i^\mu $ transforms correctly under $J^{\mu \nu}$. Let us now see how $p_i^\mu $ fares under $J^{\mu \nu}$. After a similar calculation we find,
\begin{equation}
\{J^{\mu \nu},p_1^\sigma\}=(\eta^{\mu\sigma} p_1^\nu-\eta^{\nu\sigma} p_1^\mu)~,~~\{J^{\mu \nu},p_2^\sigma \}=(\eta^{\mu\sigma} p_2^\nu-\eta^{\nu\sigma} p_2^\mu).
\end{equation}
Thus both the sets of  $x^\mu$, $p^\mu$ transform correctly under $J^{\mu \nu}$. This shows that we have been able to construct the Lorentz generator $J^{\mu \nu}$ for the full interacting theory. Furthermore, this also ensures that the Lorentz algebra $\{J^{\mu\nu },J^{\alpha\beta }\}$ of (\ref{0s}) will remain intact:
$$ \{J^{\mu\nu },J^{\alpha\beta }\} = g^{\mu\beta }J^{\nu\alpha}+g^{\mu\alpha }J^{\beta \nu}+g^{\nu\beta }J^{\alpha\mu}+g^{\nu\alpha }J^{\mu\beta}.$$

Now we discuss the Translation generator. From the set of Dirac brackets (\ref{dir1},\ref{dir2},\ref{dir3}) it is very easy to see that
\begin{eqnarray}
\{p_1^\mu +p_2^\mu,x_1^\nu\} &=& -\eta^{\mu\nu}+\frac{2}{\Lambda}p_2^2 p_1^\nu(p_1^\mu-p_2^\mu)~,~~
\{p_1^\mu +p_2^\mu ,x_2^\nu\}=-\eta^{\mu\nu}+\frac{2}{\Lambda}p_1^2 p_2^\nu(p_1^\mu-p_2^\mu)\nonumber\\
\{p_1^\mu +p_2^\mu,p_1^\nu \} &=& -\frac{\alpha^4}{4\Lambda r^6}r^\nu(p_1^\mu-p_2^\mu)~,~~
\{p_1^\mu +p_2^\mu ,p_1^\nu\}=\frac{\alpha^4}{4\Lambda r^6}r^\nu(p_1^\mu-p_2^\mu)
\label{tr}
\end{eqnarray}
Surprisingly these brackets show up new restriction for translation generator. The translation operator $T^\mu =p_1^\mu +p_2^\mu$ can act as a proper translation generator if there is a rigidity in the momentum transformation rule: $p_1^\mu -p_2^\mu =0$. This is a consequence of the fact that the interaction depends on the coordinate difference $r^\mu =x_1^\mu - x_2^\mu $. As we will show this feature will remain throughout our analysis. We will comment on this point at the end of the section. On the other hand it is reassuring to note that the translation generator algebra is correctly reproduced:
\begin{equation}
\{T^\mu,T^\nu\}=\{p_1^\mu +p_2^\mu,p_1^\nu +p_2^\nu\}=0.
\label{tr1}
\end{equation}
So far we have correctly constructed the Poincaré generators $J^{\mu\nu}$, $T^\mu$. We now focus on the dilatation generator $D$. It is straightforward to compute,

\begin{equation}
\{(x_1 p_1),x_1^\nu\}=-x_1^\nu-\frac{\alpha^4 (x_1r)}{4 \Lambda r^6}x_1^\nu+\frac{2 p_1^2 p_2^2}{\Lambda}x_1^\nu~,~~
\{(x_2 p_2),x_1^\nu\}=\frac{\alpha^4 (x_2r)}{4 \Lambda r^6}x_1^\nu+\frac{2 p_1^2 p_2^2}{\Lambda}x_1^\nu .
\label{d}
\end{equation}
Adding the  above two relations we find,
\begin{equation}
\{(x_1p_1)+(x_2p_2),x_1^\nu\}=-x_1^\nu+\frac{4 x_1^\nu}{\Lambda}\chi_3 \simeq -x_1^\nu,
\end{equation}
and a similar result for $x_2^\mu $ follows:
\begin{equation}
\{(x_1p_1)+(x_2p_2),x_2^\nu \}\mathbf{\simeq} -x_2^\nu .
\label{}
\end{equation}
Hence $D=-(x_1p_1)-(x_2p_2)$ acts as Dilation generator for the coordinates. On the other hand, when acted on  momentum variables, we find,
\begin{eqnarray}
\{(x_1p_1)+(x_2p_2),p_1^\nu\} &=&  p_1^\nu-\frac{4 p_1^2}{\Lambda}\left(p_1^2 p_2^2-\frac{\alpha^4}{16 r^4}\right)p_1^\nu=p_1^\nu -\frac{4 p_1^2}{\Lambda}p_1^\nu \chi _3 \simeq p_1^\nu,\nonumber\\
\{(x_1p_1)+(x_2p_2),p_2^\nu \} &=&  p_2^\nu+\frac{4 p_2^2}{\Lambda}\left(p_1^2 p_2^2-\frac{\alpha^4}{16 r^4}\right)p_2^\nu=p_2^\nu
+\frac{4 p_2^2}{\Lambda}p_2^\nu \chi _3 \simeq p_2^\nu .
\end{eqnarray}
This shows that $D=-(x_1p_1)-(x_2p_2)$ serves as the correct dilatation generator for momenta as well. It is also easy to check that
\begin{equation}
\left[D,D\right]=0.
\label{dd1}
\end{equation}
The remaining generator is $K^\mu $ for special conformal transformation.

Let us consider the bracket between $K^\mu $ and the translation $T^\mu$. We get,
\begin{eqnarray}
\{K_1^\mu,T^\nu\} &=& 2\left[(x_1^\mu p_1^\nu-x_1^\nu p_1^\mu)+(x_1p_1)\eta^{\mu\nu}+ (p_2^\nu-p_1^\nu)\left\{\left(\frac{\alpha^4(x_1r)}{4\Lambda r^6}+\frac{2 p_1^2 p_2^2}{\Lambda}\right)x_1^\mu+\frac{\alpha^4 x_1^2}{8\Lambda r^6}r^\mu\right\}\right]\nonumber\\
\{K_2^\mu,T^\nu\} &=& 2\left[(x_2^\mu p_2^\nu-x_2^\nu p_2^\mu)+(x_2p_2)\eta^{\mu\nu}+ (p_2^\nu-p_1^\nu)\left\{\left(\frac{\alpha^4(x_2r)}{4\Lambda r^6}+\frac{2 p_1^2 p_2^2}{\Lambda}\right)x_2^\mu-\frac{\alpha^4 x_2^2 }{8\Lambda r^6}r^\mu\right\}\right].\nonumber\\
\end{eqnarray}
Therefore adding these two brackets if $K^\mu=K_1^\mu+K_2^\mu$ we have
\begin{eqnarray}
\{K^\mu,T^\nu\} &=& 2(J_1^{\mu\nu}+J_2^{\mu\nu})+2\eta^{\mu\nu}((x_1p_1)+(x_2p_2))\nonumber\\
&+& 2(p_2^\nu-p_1^\nu) \left\{\frac{\alpha^4((x_1r)x_1^\mu+(x_2r)x_2^\mu)}{4\Lambda r^6}+\frac{2 p_1^2 p_2^2(x_1^\mu+x_2^\mu)}{\Lambda}+\frac{\alpha^4 (x_1^2-x_2^2)}{8\Lambda r^6}r^\mu\right\}
\end{eqnarray}
So by considering the special conformal transformation generator as $K^\mu =K_1^\mu+K_2^\mu$ one finds,
\begin{equation}
\{K^\mu ,T^\nu\}=2J^{\mu\nu}-2D\eta^{\mu\nu},
\label{kt}
\end{equation}
where as before the rigidity condition comes in to play to remove the $(p_1^\mu-p_2^\mu)$ terms. Also note that modulo the rigidity condition, the momenta transform correctly under $K^\mu $. This will be used later.
Moreover, with the dilatation generator $D$, the special conformal transformation generator $K^\mu$ gives the algebra
\begin{equation}
\{D,K_1^\mu\}= K_1^\mu,~\{D,K_2^\mu\}=K_2^\mu~,~~\Longrightarrow\{D,K^\mu\}= K^\mu.
\end{equation}
What remains is to compute $\{K^\mu,K^\nu \}$. We have already shown that individually the variables $x_i^\mu$, $p_i^\mu$ are transform correctly under $J^{\mu\nu }$. Then it is obvious that  the algebra between $J^{\mu\nu }$ and all the generators will be correctly reproduced, as the generators are constructed with these variables. In a similar way, to check the $\{K^\mu,K^\nu\}$ algebra we compute individually the action of $K^\mu $ on $x^\mu $ and $p^\mu $.  The algebra between $K^\mu$ and $p^\nu$ have already been produced correctly. Therefore we have to obtain algebra between $K^\mu$ and $x^\nu$:
\begin{eqnarray}
\{K^\mu,x_1^\nu\} &=& \eta^{\mu\nu} x_1^2-2x_1^\mu x_1^\nu+\frac{4p_1^2 p_2^2 x_1^\nu}{\Lambda}(x_1^\mu+x_2^\mu)+\frac{\alpha^4 x_1^\nu}{2\Lambda r^6}((x_2r)x_2^\mu-(x_1r)x_1^\mu)\nonumber\\
&&+\frac{\alpha^4 x_1^\nu r^\mu}{4\Lambda r^6}(x_1^2-x_2^2)+\frac{2 p_2^2 p_1^\nu}{\Lambda}(x_1^2 p_1^\mu-x_2^2 p_2^\mu)\nonumber\\
\{K^\mu,x_2^\nu\} &=& \eta^{\mu\nu} x_2^2-2 x_2^\mu x_2^\nu-\frac{4p_1^2 p_2^2 x_2^\nu}{\Lambda}(x_1^\mu+x_2^\mu)-\frac{\alpha^4 x_2^\nu}{2\Lambda r^6}((x_2r)x_2^\mu-(x_1r)x_1^\mu)\nonumber\\
&&-\frac{\alpha^4 x_2^\nu r^\mu}{4\Lambda r^6}(x_1^2-x_2^2)+\frac{2 p_1^2 p_2^\nu}{\Lambda}(x_1^2 p_1^\mu-x_2^2 p_2^\mu)
\end{eqnarray}
From these expressions it is clear that the algebra between $K^\mu$ and $x^\nu$ will be correctly satisfied for $x_1=-x_2$. This restriction tends to reduce the model to a Noncommutative particle with a form of self interaction.  Hence together with the rigidity condition the correct $K^\mu $ algebra will be reproduced. \\

This constitutes verification of the full conformal algebra for the Snyder two-particle interacting model in Hamiltonian formulation modulo the rigidity condition, to which we now return. One way to recover the conformal algebra is to introduce the rigidity condition mentioned earlier, $\Phi^\mu \equiv p_1^\mu-p_2^\mu \approx 0$, as a set of constraints. Using the Dirac brackets  obtained in Section 4, we  compute its bracket with the remaining First Class Constraint  $D=(x_1p_1)+(x_2p_2)$ to find,
\begin{equation}
\{D,\Phi^\mu\} = p_1^\mu -p_2^\mu-\frac{4}{\Lambda}(p_1^\mu +p_2^\mu )\left(p_1^2 p_2^2-\frac{\alpha^4}{16 r^4}\right)= \Phi^\mu . \label{D-L-bra}
\end{equation}
Hence the Conformal algebra is recovered modulo First Class Constraints. It would be interesting to explore the possibility of constructing a new set of canonical  (Darboux like) variables or modify the canonical expressions of the conformal generators so that the conformal algebra is reproduced. We hope to pursue this problem in near future. Finally if these options do not materialize, the  interacting two-particle Snyder model might be considered as a non-trivial example of a classical dynamical system where the Lagrangian symmetries are not fully manifested in the Hamiltonian framework.
\section{Conclusions and future prospects}
The motivation of studying Noncommutative spacetimes with non-constant operatorial extensions is that, they have their origin in possible Quantum Gravity scenarios and one can avoid paradoxical situations in Black Hole physics \cite{dfr}. In the absence of a fundamental theory of Quantum Gravity, the phenomenological model building is sometimes based on a top-down approach. Here one  assumes that the Quantum Gravity effects induce a form of noncommutative space-time advocated by String Theory (maybe), and then subsequently looks for signatures of Quantum Gravity  (high energy) phenomena at low energy regimes. One such NC structure is the Snyder NC spacetime proposed long ago. Conformal invariance of relativistic particle models based on Snyder spacetime has also been studied \cite{rb}. The novelty in our work is construction of a conformally invariant  Snyder-type noncommutative two particle system, with non-trivial two-body interaction. Our work is primarily based on a recent paper \cite{cas}, that discusses a two-body interacting system in conventional spacetime. Hence, in some sense we have successfully combined and extended the works of the two papers \cite{cas,rb}.

We have studied the symmetry properties both in Lagrangian and Hamiltonian framework. In the former, we deal with the explicit construction of the conformal Killing vectors. We take our cue from \cite{cas}, but once again due to the extended nature of our Snyder particle model (in comparison with the conventional particle model \cite{cas}), compatibility with the conformal killing equation further restricts the structure of the conformal killing vectors. Finally, we demonstrate that the two (apparently) independent Killing vectors, we started with, reduce to a single Killing vector.

In the Hamiltonian formalism, following Dirac's scheme of constraint dynamics, we show that the two-body system is gauge invariant (with a First Class Constraint) and lives in a coupled noncommutative spacetime, based on an extended form of Snyder algebra (by virtue of a pair of Second Class Constraints). To the best of our knowledge, this type of coupled form of noncommutative Snyder algebra has not appeared in the literature. Interesting aspects of the algebra of the conformal generators are revealed. The Lorentz generators of the interacting system are constructed that satisfy undeformed Lorentz algebra. As for the rest of the generators, they reproduce correct transformation and algebra in a restricted way. The implications of this have been discussed in the paper.

The present analysis indicates that constrained systems can show qualitatively different characteristics in Lagrangian and Hamiltonian framework, even in classical analysis (that is not taking in to account operator ordering and other quantum features). It seems that, the symmetry present in the Lagrangian framework can be manifested in a restricted way in the Hamiltonian setup. This can have interesting consequences upon quantization, and is reminiscent of the well known quantum anomaly phenomena, where not all classical symmetries are maintained upon quantization. On the other hand, it is also possible that one might be able to deform structures of the translation and special conformal transformation generators such that they satisfy undeformed algebra.

The future prospect of the present work indeed looks very promising. As we have shown in our work, although the two-body phase space admits a coupled Snyder geometry, the conformal symmetry transformations in the Lagrangian framework turn out to be structurally similar to the conventional two-body phase space, studied in \cite{cas}. Hence extension of the present noncommutative model to higher spin symmetries appears to be straightforward, at least in the Lagrangian framework. However, the resulting symplectic structure in the Hamiltonian framework will be very involved and can throw up new surprises. Another interesting aspect is the study of of dynamics, which in the present reparametrization invariant theory will be generated by the constraints. Finally, non-relativistic limit of these interacting models might be relevant in areas of recent interest, such as conformal non-relativistic hydrodynamical models \cite{con}.

\end{document}